\documentclass{ws-ijmpe}
\usepackage[super,compress]{cite}
\usepackage[breaklinks]{hyperref}
\hypersetup{colorlinks,urlcolor=black,citecolor=black,linkcolor=black,filecolor=black}
\usepackage{breakurl}
\usepackage{braket}

\begin{document}

\markboth{M. M. Saez, M. E. Mosquera, O. Civitarese}{Neutrino interactions in liquid scintillators}

\catchline{}{}{}{}{}

\title{Neutrino interactions in liquid scintillators including active-sterile neutrino mixing}

\author{M. M. Saez}

\address{Facultad de Ciencias Astron\'omicas y Geof\'{\i}sicas, University of La Plata,\\ Paseo del Bosque S/N
1900, La Plata, Argentina.\\
msaez@fcaglp.unlp.edu.ar}

\author{M. E. Mosquera}

\address{Dept. of Physics, University of La Plata, \\ c.c.~67
 1900, La Plata, Argentina.\\Facultad de Ciencias Astron\'omicas y Geof\'{\i}sicas, University of La Plata,\\ Paseo del Bosque S/N
1900, La Plata, Argentina.\\
mmosquera@fcaglp.unlp.edu.ar}

\author{O. Civitarese}

\address{Dept. of Physics, University of La Plata, \\ c.c.~67
 1900, La Plata, Argentina.\\
osvaldo.civitarese@fisica.unlp.edu.ar}

\maketitle

\begin{history}
\received{Day Month Year}
\revised{Day Month Year}
\end{history}

\begin{abstract}
Neutrinos play an important role in core-collapse supernova events since they are a key piece to understanding the explosion mechanisms. The analysis of the neutrino fluxes can bring answers to neutrino's related problems e.g.: mass hierarchy, spectral splitting, sterile neutrinos, etc. In this work we study the impact of neutrino oscillations and the possible existence of eV sterile neutrinos upon the supernova neutrino flux ($F_\nu(E)$). We have calculated the energy distribution of the neutrino flux from a supernova and the total number of events that would be detected in a liquid scintillator. We also present an analysis for the conversion probabilities as a function of the active-sterile neutrino mixing parameters. Finally, we have carried out a statistical analysis to extract values for the mixing parameters of the model.
\end{abstract}


\ccode{PACS numbers:}

\section{Introduction}

Core-collapse supernovae (SN) represents the final evolutionary stage of stars with masses heavier than 8 $M_\odot$. To explain these events one needs to combine nuclear and particle physics with astrophysics. Neutrinos are important to the study of the energy balance involved in SN collapses, since only about $1\%$ of the gravitational binding energy is released as kinetic energy in the compact object formation, while the remaining $99\%$ is carried out by neutrinos with energies of several MeV \cite{Woosley:1986}. In the SN core, a medium of high temperature and density, neutrinos of all flavors are produced. The neutrino production mechanisms are, mainly, pair annihilation  $e^+ + e^- \rightarrow \nu_e+\bar{\nu_e}$, flavor-conversion $\nu_e+\bar{\nu_e}\rightarrow \nu_{\tau,\mu}+\bar{\nu}_{\tau,\mu}$, and  nucleon bremsstrahlung $N+N\rightarrow N+N+\nu_{\tau,\mu}+\bar{\nu}_{\tau,\mu}$ \cite{Buras:2003}. Neutrinos, that travel through stellar material and space, can reach a neutrino detector on Earth, giving information from deep inside the stellar core. During the accretion phase, which takes place few tens to hundreds of milliseconds after the bounce, the expected neutrino energy spectrum would exhibit a flavor hierarchy $\braket{E_{\nu_e}}< \braket{E_{\bar{\nu}_e}}<\braket{E_{\nu_x}}$ \cite{Janka:2016}. However,  the presence of neutrino oscillations can alter the composition of the flux that reaches the Earth \cite{Balantekin:2005,Tamborra:2012,Rafflet:2007,Dasgupta:2009}.  

Neutrino oscillations have been studied during the last decades, and different detectors on Earth have measured neutrino fluxes generated by reactions which take place in the sun, the Earth atmosphere, reactors, and supernovae like the SN 1987A \cite{Giunti:2003,Abe:2016,Gallex:1998,Kajita:2010,Lunardini:2001,Hirata:1988}. The analysis and reconstruction of SN neutrino fluxes is a powerful tool to clarify the role of neutrinos in stellar explosions and nucleosynthesis \cite{Qian:2018}, as well as studying physics beyond the standard model \cite{Mirizzi:2016,Takahashi:2001,Balantekin:2005, Machado:2012}. 

The possibility of the existence of a light sterile neutrino, motivated by some experimental anomalies detected in short-baseline neutrinos oscillation experiments \cite{Aguilar-Arevalo:2018,Dentler:2018}, in reactor experiments \cite{Mention11} and gallium detectors \cite{Acero:2007,Giunti:2011}, is currently under investigation. This extra neutrino does not participate in weak interactions, and interacts only  gravitationally, thus it can participate in the mixing processes with active neutrinos \cite{Kopp:2013}. The consequences of the existence of sterile neutrinos in different astrophysical scenarios have being examined in previous works \cite{Boyarsky:2009,Mohapatra:2004}, among others. In particular, in the context of SN, several authors have already analysed the effects of the inclusion of a sterile neutrino upon the fraction of free neutrons, the baryonic density, the electron fraction of the material, and nucleosynthesis processes \cite{Mclaughlin:1999, Caldwell:2000,Fetter:2003,Qian:2018,Saez:2018,Tamborra:2012,Wu:2014,Esmaili:2014}. In particular, in reference \cite{Tang:2020}, the SN events produced via proton and electron elastic scattering in scintillators are studied.

The effects due to the transformation between active and sterile neutrino species have been advanced, with reference to r-process nucleosynthesis, in the paper of G.C.Mc Laughlin, J.M.Fetter, A.B. Balantekin and G.M.Fuller  \cite{Mclaughlin:1999}.  In this work it is shown that the mixing between sterile and active electron-neutrinos  could enhance the rapid neutron capture in supernovae, and limits are set for the square-mass-difference and for mixing angle. 

The formalism of  \cite{Mclaughlin:1999} is based on the solution of a two by two time-dependent Schroedinger equation where diagonal terms are density dependent and non-diagonal terms are just the vacuum square-mass differences between actrive and sterile neutrino species. The theoretical framework applies both to electron-neutrinos and to electron-antineutrinos in their mixing with sterile ones.

In the present work we have taken most of the basic elements  presented in Ref.\cite{Mclaughlin:1999} as 
a motivation for our study and  esentially followed the same type of arguments to compute neutrino fluxes in dealing with charged current induced reactions like the inverse beta decay.

{\bf{We have choosen this channel (charge currents with electron antineutrinos) in view of its relatively large contribution to the cross section, as reported by the SNO+ \cite{SNOplus:2015} collaboration. It amounts to nearly half the value (approx 200 events) of the neutral current events mediated by electron neutrinos (approx 400 events). 
Charged current (CC) antineutrino (neutrino) reactions on C leading to the ground state of Be and N amount to a small but perhaps not still negligible fraction of the total cross section (e.g: about 7 events for Be and 3 events for N to be compared to approx 195 events for the electron-antineutrino + p reactions \cite{SNOplus:2015}).
However, the impact of the mixing between sterile and active neutrinos upon  other reactions, like neutral and charged currents on Carbon isotopes, may not be negligleable. The estimated number of events for these reactions can be about 10 times smaller, as shown in the same table of events for SNO+, but still they are of interest. A very complete compilation of results of calculations for neutral and charged currents, covering an extended domain of values for the neutrino (antineutrino) energy in its interaction with $^{12,13}$C, has been presented in Refs.\cite{Suzuki:2006,Suzuki:2019}.}}

In this work, we focus on the study of SN neutrino signals produced by inverse beta decays in a liquid scintillator and how they are affected by the inclusion of the oscillations and by the inclusion of a light sterile neutrino in the formalism. We have analyzed the effects of different mixing parameters in the $3+1$ scheme, upon some observables as well as the possibility of having sterile neutrinos in the initial composition, a possibility not addressed in previous works such as \cite{Tang:2020,Choubey:2006}. In addition, we have studied the adiabaticity of the transition between active and sterile $\nu$-species as a function of the mixing parameters and performed a statistical analysis in energy bins to find values for the unknown parameters of the model.

This paper is organized as follows. In Sec. \ref{formalismo} we introduce the formalism needed to calculate SN neutrino fluxes and the neutrino interactions in liquid scintillators. In Sec. \ref{resultados}, we present and discuss the results of the calculation of the neutrinos fluxes, number of events and crossing probabilities when active-active  and active-sterile neutrino oscillations are included in the formalism. The conclusions are drawn in Sec. \ref{conclusiones}.

\section{Formalism}
\label{formalismo}

\subsection{Neutrino fluxes and crossing probabilities}

We follow \cite{Huang:2015} and consider the standard energy released by the supernova (SN) neutrino outflow similar to the SN1987A \cite{Hirata:1987,Bionta:1987}, that is $E_\nu^{tot}=3\times 10^{53} \, {\rm erg}$ distributed between different neutrino's flavors. The luminosity flux, for each flavor, is time dependent and can be written as
\begin{equation}\label{eq:lum}
L_{\nu_\beta}(t)=\frac{E_\nu^{tot}}{18} e^{-t/3} \, \, \, .
\end{equation}

From the previous expression one can write the spectral flux for neutrinos of each flavor which are produced in the SN explosion and which are detected at a distance $D$ from the SN, in units of $\rm{MeV}^{-1} \rm{cm}^{-2}$, as
\begin{equation}\label{eq:FD}
F^0_{\nu_\beta}(E,\, t)=\frac{L_{\nu_\beta}(t)}{4\pi D^2} \frac{ f_{\nu_\beta}(E,\eta)}{\braket{E_{\nu_\beta}}}\,\,\, ,
\end{equation}
where $f_{\nu_\beta}$ is the neutrino distribution function, $E$ is the neutrino energy and $\braket{E_{\nu_\beta}}$ is the mean energy of the $\beta$-flavor neutrino eigenstate. 

There are two different distribution functions used in the computation of SN neutrino spectrum, i) the Fermi-Dirac distribution 
\begin{equation}\label{eq:Fermi}
f_{\nu_\beta}(E, \, \eta_{\nu_\beta})=\frac{1}{1+exp[E/T_{\nu_\beta}-\eta_{\nu_\beta}]} \, \, , 
\end{equation}
where the neutrino temperature is related with its mean energy as $\braket{E_{\nu_\beta}}\propto T_{\nu_\beta}$ if $\eta_{\nu_\beta}=0$ and $\braket{E_{\nu_\beta}}\propto T^2_{\nu_\beta}$ if $\eta_{\nu_\beta}\neq0$, being $\eta_{\nu_\beta}$  the pinching parameter; and ii) the power law distribution \cite{Keil:2003}
\begin{equation}\label{eq:PL}
f_{\nu_\beta}(E)= \frac{(\alpha+1)^{(\alpha+1)}}{\Gamma(\alpha+1)\braket{E}} \left(\frac{E}{\braket{E}}\right)^\alpha e^{-(\alpha+1)E/\braket{E}}\, \, .
\end{equation}
In the previous equations both distribution functions are normalized, therefore $F_2$ is the Fermi integral of order 2 and $\braket{E^2}/\braket{E}^2=(\alpha+2)/(\alpha+1)$.

If the neutrino flux is thermalized one can used a Fermi Dirac distribution function with $\eta=0$, however, in a SN neutrino flux this condition in not always fulfilled. Therefore one uses the power law distribution function instead \cite{Tamborra:2012,Keil:2003,Janka:1989,Lang:2016,Huang:2015}.

\subsubsection{Time evolution and mixing scheme}

The time evolution of neutrinos interacting with electrons, in the flavor basis, is given by the equation 
\begin{equation}\label{evolution}
i \frac{d}{dt}{\psi_{\nu_\beta}}=(H_{vac}+V)\psi_{\nu_\beta} \, \, ,
\end{equation}
where $\psi_{\nu_\beta}$ is the neutrino wave function, $V$ represents the interactions between neutrinos and electrons, $V= \sqrt{2} G_F {\rm diag}(N_e,0,0)$, $G_F$ is the Fermi constant and $N_e$ is the electron number density \cite{Tang:2020}. For antineutrinos, the potential has opposite sign. Near the SN core, the neutrino density is so high that the neutrino-self interactions may affect the flavor evolution in a non-trivial way \cite{Duan:2010b}. These effects are only
partially understood and have been modelled under several simplifications so results concerning those effects have been neglected in the following. Also, for the early time signal during the neutronization burst, such effects are almost absent, and concerning the accretion phase, they are found to be suppressed by multi-angle “matter” effects \cite{Chakraborty:2011, Esteban:2008}. The vacuum Hamiltonian $H_{vac}$ in the flavor basis can be obtained from the mass Hamiltonian as 
\begin{equation} \label{H}
H_{vac}=\frac{U M^2 U^{\dagger}}{2E} \, \, \, .
\end{equation}
In the last expression $M^2$ is the matrix of square mass differences, $M^2={\rm diag}\left(0, \, \Delta m^2_{21}, \, \Delta m^2_{31}\right)$ for the $3$-scheme, and $M^2={\rm diag}\left(0,\, \Delta m^2_{21}, \, \Delta m^2_{31}, \, \Delta m^2_{41}\right)$ for the $3+1$-scheme, with the standard notation $\Delta m^2_{ij}=m^2_i-m^2_j$, being $m_j$ the mass of the neutrino in the $j$-eigenstate. The unitary matrix $U$ is the PMNS mixing matrix,  \cite{Maki:1962,Giganti:2018}, which for the standard $3$-mass scheme is written
\begin{equation}\label{u3}
U=\left(
\begin{array}{ccc}
c_{12} c_{13} & s_{12} c_{13} & s_{13} \\
\alpha & \beta & s_{23} c_{13} \\
\delta  & \omega & c_{23} c_{13}
\end{array}
\right) \, \, \, ,
\end{equation}
where $c_{ij}$ $\left(s_{ij}\right)$ stands for $\cos \theta_{ij}$ $\left(\sin \theta_{ij}\right)$ (i,j=1,2,3), and $\alpha= -s_{12} c_{23} -c_{12} s_{23} s_{13}$, $\beta =c_{12} c_{23} -s_{12} s_{23} s_{13}$, $\delta= s_{12} s_{23} -c_{12} c_{23} s_{13}$ and $\omega = -c_{12} s_{23} -s_{12} c_{23} s_{13}$.

The  U-matrix for the 3+1 scheme has the form \cite{Gariazzo:2017,Collin:2016}
\begin{equation}\label{U-3+1}
U=
\left(
\begin{array}{cccc}
c_{12} c_{13} c_{14} & s_{12} c_{13} c_{14} & s_{13} c_{14} & s_{14}\\
\alpha & \beta & s_{23} c_{13} &0\\
\delta & \omega  & c_{23 }c_{13} &0 \\
-c_{12} c_{13} s_{14} & -s_{12} c_{13} s_{14} & -s_{13} s_{14} & c_{14}
\end{array}
\right) \, \, \, ,
\end{equation}
where the parameter $\theta_{14}$ is added in order to account for the mixing between the sterile and the lightest neutrino mass eigenstate.

\begin{figure*}[ht]
\centering
\includegraphics[width=0.85\textwidth]{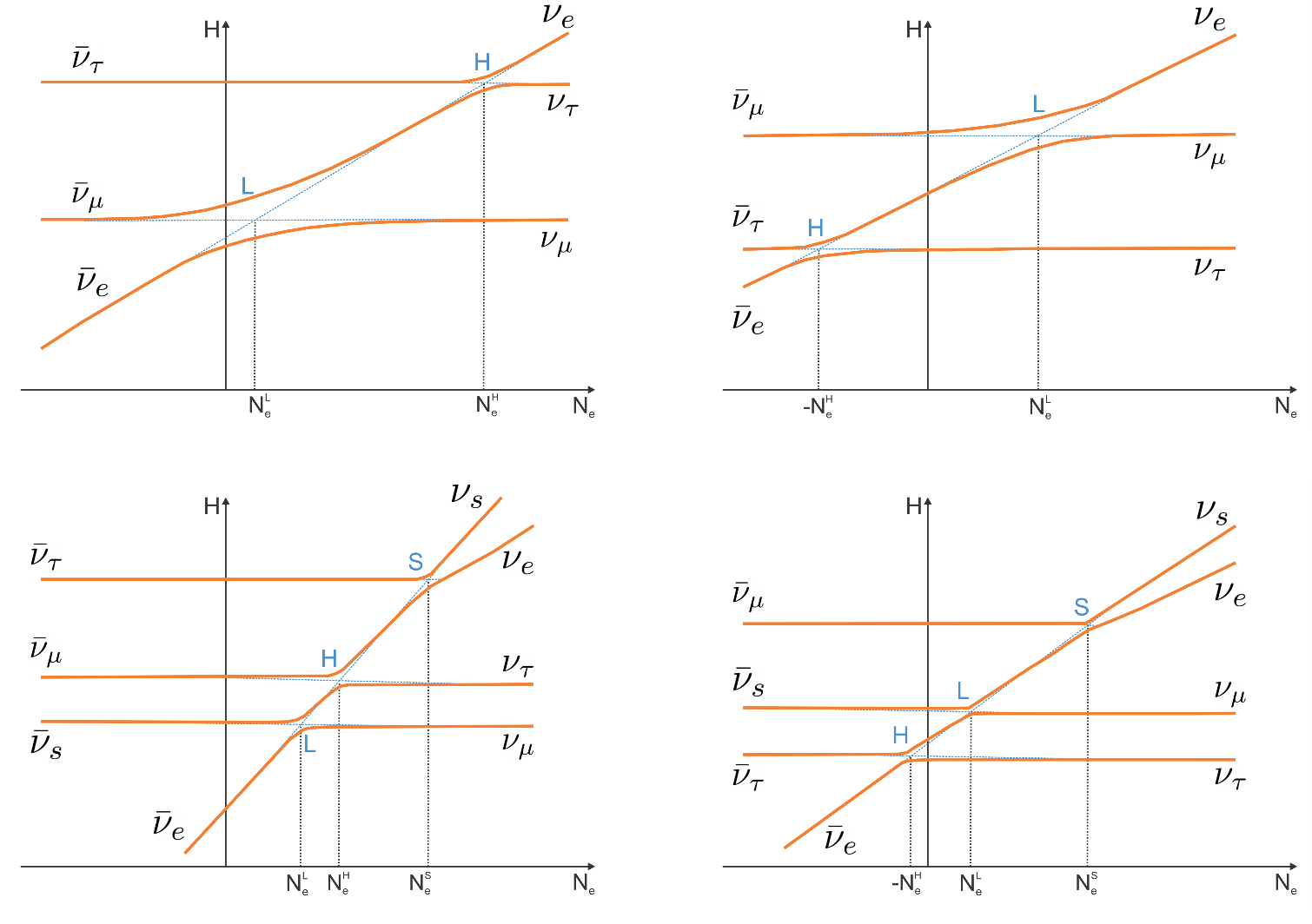}
\caption{Crossing diagram. Left hand-side: normal hierarchy; right hand-side: inverted hierarchy. Top panel: $3$ active neutrinos; bottom panel: $3+1$ neutrino scheme.}\label{crossing}
\end{figure*}

In the two schemes studied, we have taken the mixing mechanism to be unaffected by CP violations. In both cases, one can perform a rotation in this subspace to diagonalize the submatrix $\mu \, \tau$ of Eq.(\ref{H}) \cite{Dighe:2000,Tamborra:2012}, and since the $\nu_\mu$ and $\nu_\tau$ fluxes in the SN are similar we denote them by $F_{\nu_x}$. Neutrino flavor conversions inside a supernova are possible and their effectiveness depends on the matter density. Since the matter density (and therefore the potencial) decreases with the star radius, active neutrinos exhibit two Mikheyev-Smirnov-Wolfenstein (MSW) resonances called H (high density) and L (low density) where the flavor conversion mechanism is amplified \cite{Dighe:2000}. One can solve Eq. (\ref{evolution}) and plot the neutrino level crossing scheme of Fig. \ref{crossing} (the half-plane with positive values of density corresponds to neutrinos and the half-plane with negative values of the density to antineutrinos) \cite{Dighe:2000,Giunti:2003,Smirnov:2003}. If we consider sterile neutrinos, there is an inner extra resonance, the S resonance, between the electron type neutrino and the sterile one. It is worth mentioning that in other works, the resonance is considered to occur at low densities (outer resonance) \cite{Tang:2020}

The neutrino flux that arrives at Earth can be computed as a superposition of the neutrino fluxes produced in the neutrinosphere inside the SN. 

\subsubsection{Neutrino fluxes in the 3-active scheme}

Following the top row diagrams of Fig. \ref{crossing}, the flux for
$\nu_e$ and $\nu_x$ neutrinos (anti-neutrinos) is written:
\begin{eqnarray}
F_{\nu_e}&=& P_{e} F^0_{\nu_e} + (1-P_e)F^0_{\nu_x} \, \, \, , \nonumber \\
F_{\bar{\nu}_e}&=& \bar{P}_e F^0_{\bar{\nu}_e} + (1-\bar{P}_e)F^0_{\bar{\nu}_x} \, \, \, , \nonumber \\
F_{\nu_x}&=& (1-P_e) F^0_{\nu_e} + (1+P_e)F^0_{\nu_x}\, \, \, , \nonumber \\
\label{flujos-3}
F_{\bar{\nu}_x}&=& (1-\bar{P}_e) F^0_{\bar{\nu}_e} + (1+\bar{P}_e)F^0_{\bar{\nu}_x}\, \, \, .
\end{eqnarray}
In the previous equations $P_e$ and $\bar{P}_e$ are the survival probabilities for $\nu_e$ and ${\bar{\nu}_e}$ respectively, which depend on the hierarchy. That is:
\begin{eqnarray}
P_e&=&|U_{e1}|^2 P_H P_L +|U_{e2}|^2 P_H (1-P_L) \nonumber \\
&& + |U_{e3}|^2(1-P_H) \, \, \, , \nonumber \\
\bar{P}_e&=& |U_{e1}|^2 \, \, \, ,
\end{eqnarray}
for normal hierarchy, and
\begin{eqnarray}
P_e&=&|U_{e1}|^2P_L+|U_{e2}|^2(1-P_L)  \, \, \, , \nonumber \\
\bar{P}_e&=& |U_{e1}|^2\bar{P}_H+|U_{e3}|^2(1-\bar{P}_H) \, \, \, ,
\end{eqnarray}
for inverse hierarchy, respectively.

In last equations $U_{ek}$ $(k=1,\, 2,\, 3)$ are the components of the first row of the U-matrix (see Eq.\ref{u3}). $P_H$ ($\bar{P}_H$) and $P_L$ ($\bar{P}_L$) are the neutrino (antineutrino) crossing probabilities for the H and L resonances respectively (see  Section \ref{crossing-prob}).

\subsubsection{Neutrino fluxes in the 3+1 scheme}

Following the bottom row of Fig. \ref{crossing}, the fluxes for the $3+1$ scheme are
\begin{eqnarray}
F_{\nu_e}&=& \Theta_e^e F^0_{\nu_e} + \Theta_e^x F^0_{\nu_x} + \Theta_e^s  F^0_{\nu_s}  \, \, \, , \nonumber \\
F_{\bar{\nu}_e}&=& \Xi_e^e F^0_{\bar{\nu}_e}+ \Xi_e^x F^0_{\bar{\nu}_x}+\Xi_e^s F^0_{\bar{\nu}_s}  \, \, \, , \nonumber \\
F_{\nu_x}&=& \left(\Theta_\mu^e +\Theta_\tau^e\right) F^0_{\nu_e} + \left(\Theta_\mu^x +\Theta_\tau^x\right) F^0_{\nu_x} \nonumber \\
&& + \left(\Theta_\mu^s +\Theta_\tau^s\right)  F^0_{\nu_s}\, \, \, , \nonumber\\ 
F_{\bar{\nu}_x}&=& \left(\Xi_\mu^e +\Xi_\tau^e \right) F^0_{\bar{\nu}_e}+ \left(\Xi_\mu^s +\Xi_\tau^s \right)F^0_{\bar{\nu}_x}\nonumber \\
&&+ \left(\Xi_\mu^s +\Xi_\tau^s \right) F^0_{\bar{\nu}_s}\, \, \, , \nonumber\\
F_{\nu_s}&=&\Theta_s^e F^0_{\nu_e} + \Theta_s^x F^0_{\nu_x} +\Theta_s^s F^0_{\nu_s}\, \, \, , \nonumber \\
\label{flujos-3+1}
F_{\bar{\nu}_s}&=&\Xi_s^e F^0_{\bar{\nu}_e}+\Xi_s^x  F^0_{\bar{\nu}_x}+\Xi_s^s F^0_{\bar{\nu}_s} \, \, \, ,
\end{eqnarray}
where for normal hierarchy we have defined
\begin{eqnarray}
\Theta_{\alpha}^e&=& |U_{\alpha 1}|^2P_H P_L (1-P_S)+|U_{\alpha 3}|^2P_S\nonumber \\
&& +|U_{\alpha 2}|^2 P_H(1-P_L)(1-P_S) \nonumber \\
&&+|U_{\alpha 4}|^2(1-P_S)(1-P_H) \, \, \, ,\nonumber\\
\Theta_{\alpha}^x&=& |U_{\alpha 1}|^2(1-P_H P_L) +|U_{\alpha 2}|^2(1-P_H+P_H P_L)\nonumber \\
&&+|U_{\alpha 4}|^2P_H  ,\nonumber\\
\Theta_{\alpha}^s&=& |U_{\alpha 1}|^2P_S P_H P_L+|U_{\alpha 2}|^2P_S P_H(1-P_L) \, \, \, , \nonumber \\
&&+|U_{\alpha 3}|^2(1-P_S)+|U_{\alpha 4}|^2P_S(1-P_H)  \, \, \, , \nonumber \\
\Xi_{\alpha}^e&=&|U_{\alpha 1}|^2\, \, \, , \nonumber \\
\Xi_{\alpha}^x&=& |U_{\alpha 2}|^2+|U_{\alpha 3}|^2\, \, \, , \nonumber \\
\Xi_{\alpha}^s&=& |U_{\alpha 4}|^2\, \, \, . 
\end{eqnarray}
For the inverse hierarchy we have called
\begin{eqnarray}
\Theta_{\alpha}^e&=& |U_{\alpha 1}|^2P_L (1-P_S)+|U_{\alpha 2}|^2P_S\nonumber\\
&&+|U_{\alpha 4}|^2(1-P_S)(1-P_L) \, \, \, ,\nonumber\\
\Theta_{\alpha}^x&=& |U_{\alpha 1}|^2(1-P_L) +|U_{\alpha 3}|^2+|U_{\alpha 4}|^2 P_L  \, \, \, ,\nonumber\\
\Theta_{\alpha}^s&=&  |U_{\alpha 1}|^2P_SP_L+|U_{\alpha 2}|^2(1-P_S)\nonumber\\
&&+|U_{\alpha 4}|^2P_S(1-P_L) \, \, \, , \nonumber\\
\Xi_{\alpha}^e&=& |U_{\alpha 2}|^2\bar{P}_H+|U_{\alpha 3}|^2(1-\bar{P}_H)\, \, \, , \nonumber \\
\Xi_{\alpha}^x&=& |U_{\alpha 1}|^2+|U_{\alpha 2}|^2(1-\bar{P}_H)+|U_{\alpha 3}|^2\bar{P}_H\, \, \, , \nonumber \\
\Xi_{\alpha}^s&=& |U_{\alpha 4}|^2\, \, \, . 
\end{eqnarray}

In the last expressions $U_{\alpha k}$ $(k=1,\, 2,\, 3,\, 4)$ are the elements of the mixing matrix (see Eq.(\ref{U-3+1})) and $P_S$ stands for the probability of crossing the S-resonance. To calculate the initial sterile flux $F^0_{{\nu}_s}$ we follow Eqs.(\ref{eq:lum}-\ref{eq:PL}).

\subsubsection{Crossing probabilities}\label{crossing-prob}

As we have seen in the previous sections, the fluxes can be expressed in terms of the crossing probabilities $P_H$, $P_L$ and $P_S$. These are related to the adiabatic parameter $\gamma$ as \cite{Landau:1932,Zener:1932}
\begin{equation} \label{pc}
P= e^{-\pi \gamma /2} \, \, .
\end{equation}
The adiabatic parameter depends on the change of the density with the radius as
\begin{equation}
\gamma=\frac{\Delta m^2\sin^2(2\theta)}{2E\cos(2\theta)\left|\frac{d\ln(N_e)}{dr}\right|_{res}}\, t\, ,
\end{equation}
where $\theta$ is the neutrino mixing angle and $\Delta m^2$ is the square mass difference. In order to describe the SN environment, we have considered $N_e=\rho/m_N= A/(m_N r^3)$ where $A$ is a constant and $m_N$ is the neutron mass \cite{Mirizzi:2016,Fogli:2002,Brown:1982,Dighe:2000}. The MSW resonance condition can be written as  \cite{Mikheev:1986,Tang:2020} 
\begin{equation}
\cos(2\theta)=\frac{2 V_{ee} E}{\Delta m^2}\, \, ,
\end{equation}
therefore 
\begin{equation}
\gamma=\frac{1}{3}\left(\frac{\Delta m^2}{2E}\right)^{2/3}\frac{\sin^2(2\theta)}{\cos^{4/3}(2\theta)}\left(\frac{\sqrt{2}G_F A Y_e}{ m_N}\right)^{1/3} \, \, .
\end{equation}
If this parameter is larger than $1$, the crossing probability of Eq.(\ref{pc}) is almost zero and the regime becomes  adiabatic.

In order to compute the different probabilities associated to the three resonances, we have considered the mixing parameters of atmospheric neutrinos $(\Delta m_{31}^2$ and $\theta_{13})$ for the H resonance $(P_H)$, of the solar neutrinos  $(\Delta m_{21}^2$ and $\theta_{12})$ for the L resonance $(P_L)$ and for the probability of the S resonance $(P_S)$ we used the sterile active neutrino oscillation parameters $(\Delta m_{41}^2$ and $\theta_{14})$. The position of these resonances are shown schematically in Fig. \ref{crossing}. 

\subsection{Neutrino detection}

Once the SN neutrino burst arrives to the Earth it can be detected by several experimental arrays,  such as GALLEX/GNO \cite{Gallex:1998,Altmann:2005}, SAGE \cite{Sage:1999}, Kamiokande \cite{Nakamura:1994}, Super-Kamiokande \cite{Fukuda:2002}, MiniBooNE \cite{Aguilar-Arevalo:2018}, KamLAND \cite{Kamland:2005}, Borexino \cite{Borexino:2002} , LSND \cite{LSND:1996}, SNO+ \cite{SNOplus:2015}.

For the case of neutrino detection in liquid scintillators, such as KamLand, Borexino and SNO+, the charged current interactions between neutrinos and nuclei are the inverse beta decay $\left(\bar{\nu_e} + p \rightarrow n + e^+\right)$,  and the neutrino (antineutrino) capture on carbon  $\nu_e + ^{12}C \rightarrow e^-+^{12}N$ ($\bar{\nu}_e + ^{12}C \rightarrow e^+ +^{12}B$)\cite{Suzuki:2006}, and the similar reactions on $^{13}C$\cite{Suzuki:2019}. The first kind of interaction is dominant since its cross-section is much larger, therefore the number of events can be computed as \cite{Strumia:2003}
\begin{equation}
\label{num-eventos}
N=N_p\int_{E_{min}}^\infty dE F_{\bar{\nu}_e} (E) \sigma_{\bar{\nu}_{e}}(E) \, \, ,
\end{equation}
where $N_p$ is the number of free protons in the detector, $E$ is the energy and $E_{min}=1.806 \, \rm{MeV}$ is the threshold energy. For example, a $3\, {\rm kT}$ detector of ${\rm C}_6 {\rm H}_5 {\rm C}_{12} {\rm H}_{25}$ (SNO \cite{SNOplus:2015}) has $N_p=2.2\times 10^{32}$ free protons, the KamLAND detector, made of $80\%$ ${\rm C}_{12} {\rm H}_{26}$ and $20\%$ ${\rm C}_9 {\rm H}_{12}$, has $N_p=1.7 \times 10^{33}$ free protons \cite{Kamland:2005} and  the detector located in Borexino $\left({\rm C}_9 {\rm H}_{12} \right)$ has $N_p=1.81 \times 10^{31}$ \cite{Borexino:2002,Cadonati:2002}. As our reference detector, we adopt a $3\, {\rm kT}$ one based on alkyl benzene since it represents a suitable option for the class of detectors to be installed in the ANDES laboratory. The results presented hereinafter correspond to this choice \cite{Machado:2012,Civitarese:2015}. The cross section for the inverse beta decay $\sigma_{\bar{\nu}_{ep}}$, in units of $10^{-43}\, {\rm cm}^2$, can be approximated as \cite{Strumia:2003}
\begin{equation}\label{sig(e)}
\sigma_{\bar{\nu}_{e}}(E)=p_e E_e E^{-\,0.07056\, +\,0.02018\, y\,-\,0.001953 \, y^3} \, \, ,
\end{equation}
for energies lower than $300 \, {\rm MeV}$. In the previous equation $p_e$ stands for the positron momentum related to the neutrino energy $E$ (in MeV) as $p_e^2=(E-\Delta)^2-m_e^2$, where the neutron to proton mass difference is $\Delta=m_n-m_p=1.293 \, {\rm MeV}$ and $m_e$ is the positron mass. The neutrino and positron energy are related by $E_e=E -\Delta$, and $y= \ln E $ in the exponent of Eq.(\ref{sig(e)}).

\section{Results and Discussion}
\label{resultados}

\subsection{Active-active neutrino mixing}

In order to calculate the SN neutrino flux and the number of events in a detector, we have adopted the parameters given in Ref. \cite{Dighe:2000}, and fixed  the density at the value $A Y_e=2\times 10^{16}\, {\rm kg/km}^3$. We have performed the integration of the luminosity flux of Eq. (\ref{eq:lum}), in a time-interval of 20 seconds \cite{Hirata:1987,Bionta:1987} and the SN-detector distance was fixed at $D=10\, {\rm kpc}$ \cite{Mirizzi:2006,Machado:2012}. For the active-active neutrino oscillation parameters needed to compute $P_e$ and $\bar{P}_e$ we have used the values provided by the Particle Data Group  \cite{Pdg:2019} listed  in Table \ref{dataactivos}. 
\begin{table}[htp]
\begin{center}
\begin{tabular}{|c|c|c|}\hline
Parameter & Normal hierarchy & Inverse hierarchy \\ \hline
$\sin^2(\theta_{12})$ & $0.307$ & $0.307$\\ \hline
$\Delta m^2_{21}$ & $7.53 \times 10^{-5} \, {\rm eV}^2$ & $7.53 \times 10^{-5} \, {\rm eV}^2$\\ \hline
$\sin^2(\theta_{23})$ & $0.545$ & $0.547$\\ \hline
$\Delta m^2_{32}$ & $2.46 \times 10^{-3} \, {\rm eV}^2$ & $-2.53 \times 10^{-3} \, {\rm eV}^2$\\ \hline
$\sin^2(\theta_{13})$ & $0.0218$ & $0.0218$\\ \hline
\end{tabular}
\caption{Active-active neutrino mixing parameters \cite{Pdg:2019}.} \label{dataactivos}
\end{center}
\end{table}

To compute the SN neutrino flux which arrives to Earth (see Eq. (\ref{flujos-3})), we use the two previously introduced distribution functions as initial conditions, the Fermi-Dirac (FD) distribution of Eq. (\ref{eq:FD}),  and the power law  (PL) distribution of Eq. (\ref{eq:PL}). For the FD distribution functions we have considered three different sets of values for $\eta_{\nu_\beta}$: 
\begin{itemize}
\item [i)] FD0: $\eta_{\nu_\beta}=0$ for all the flavors; 
\item [ii)] FD1: $\eta_{\nu_e}=1.7$, $\eta_{\bar{\nu}_e}=3$ and $\eta_{\nu_x}=\eta_{\bar{\nu}_x}=0.8$ \cite{Keil:2003,Janka:1989};
\item [iii)] FD2: $\eta_{\nu_e}=3$, $\eta_{\bar{\nu}_e}=2$ and $\eta_{\nu_x}=\eta_{\bar{\nu}_x}=0.8$ \cite{Huang:2015}.
\end{itemize}
For the PL distribution we have used the value $\alpha=3$.

The mean energies used for the calculation  of neutrino fluxes are $\braket{E_{\nu_e}}=12 \, {\rm MeV}$, $\braket{E_{\bar{\nu}_e}}=15 \, {\rm MeV}$ and $\braket{E_{\nu_x}}=\braket{E_{\bar{\nu}_x}}=18 \, {\rm MeV}$ \cite{Machado:2012}. If one uses the values for the mixing parameters given  in Table \ref{dataactivos}, for the  adopted density profile, the crossing probabilities $P_H$ and $P_L$ are quite small, indicating an adiabatic crossing.

\begin{figure*}[htp]
\begin{center}
\includegraphics[width=1 \textwidth]{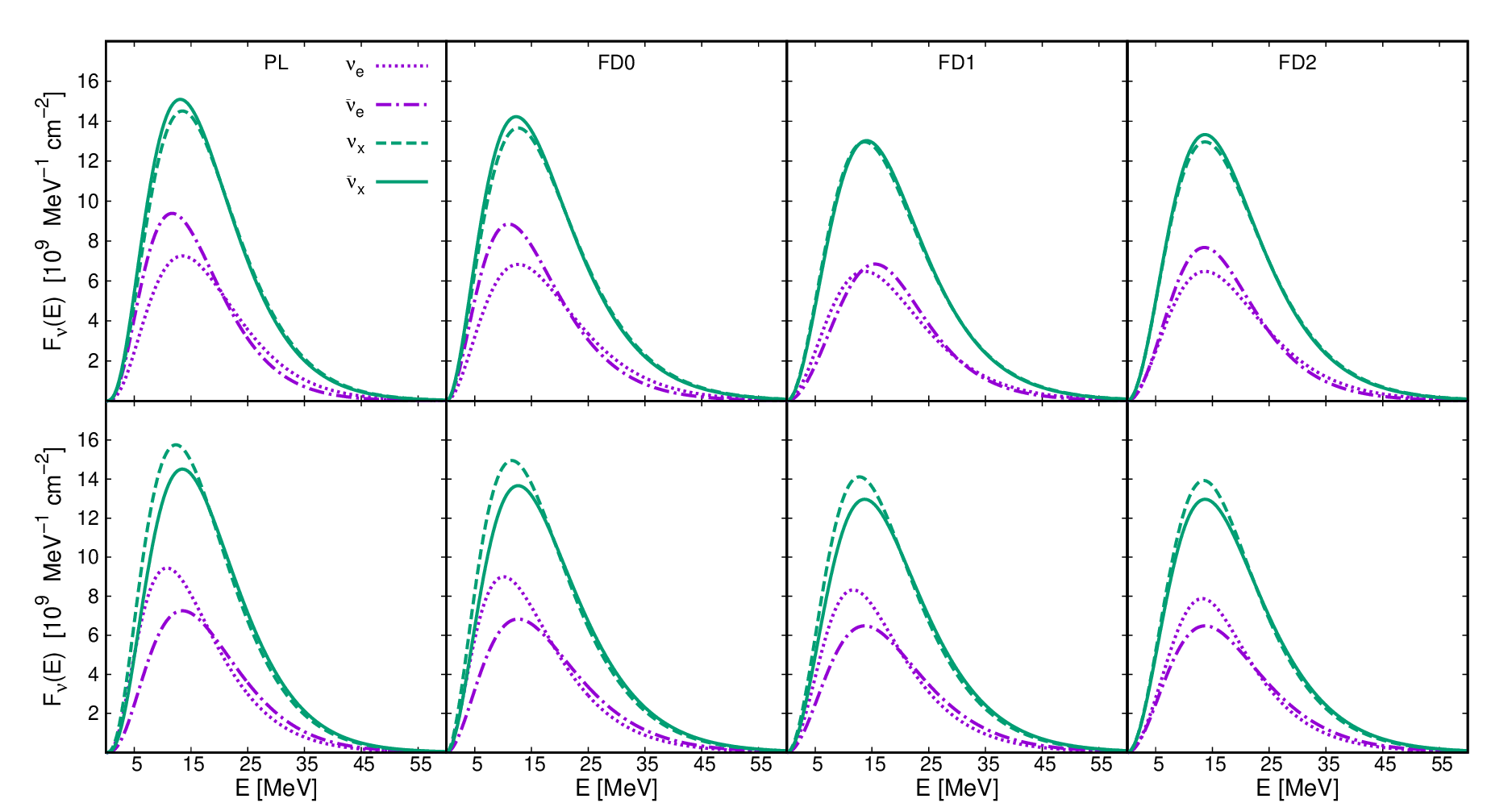}
\end{center}
\caption{Neutrino's fluxes at the detector as a function of the neutrino energy, for the active-active neutrino mixing scenario. Top row: normal hierarchy; bottom row: inverse hierarchy. Form left to right: first column: power law distribution function (PL); second column: Fermi Dirac with $\eta_{\nu_\beta}=0$ (FD0); third column: Fermi Dirac distribution function with $\eta_{\nu_e}=1.7$, $\eta_{\bar{\nu}_e}=3$ and $\eta_{\nu_x}=\eta_{\bar{\nu}_x}=0.8$ (FD1); fourth column: Fermi Dirac distribution function with $\eta_{\nu_e}=3$, $\eta_{\bar{\nu}_e}=2$ and $\eta_{\nu_x}=\eta_{\bar{\nu}_x}=0.8$ (FD2).}\label{fig:flujo-3act}
\end{figure*}

In Fig. \ref{fig:flujo-3act} we show the SN neutrino's fluxes at the detector, as a function of the neutrino energy, for different initial conditions (PL, FD0, FD1, and FD2) and for both hierarchies. The inclusion of non-zero values for the chemical potentials (that is FD1 and FD2) reduces all of the neutrino fluxes. The flux of the electron antineutrino (the one that can be detected in scintillator detectors) depends strongly upon the mass hierarchy. For normal hierarchy (NH), the electron-antineutrino flux dominates over the electron-neutrinos but for the inverse hierarchy (IH), a swap between electronic fluxes occurs. Furthermore, for IH the peaks of the electron antineutrinos fluxes locate at higher energies with respect to the case corresponding to the normal hierarchy.
If the neutrino mean energies are larger, for instance $\braket{E}=22 \, {\rm MeV}$, the neutrino fluxes decrease, as expected from other studies \cite{Keil:2003,Machado:2012}.

\begin{figure}[htp]
\begin{center}
\includegraphics[width=0.45 \textwidth]{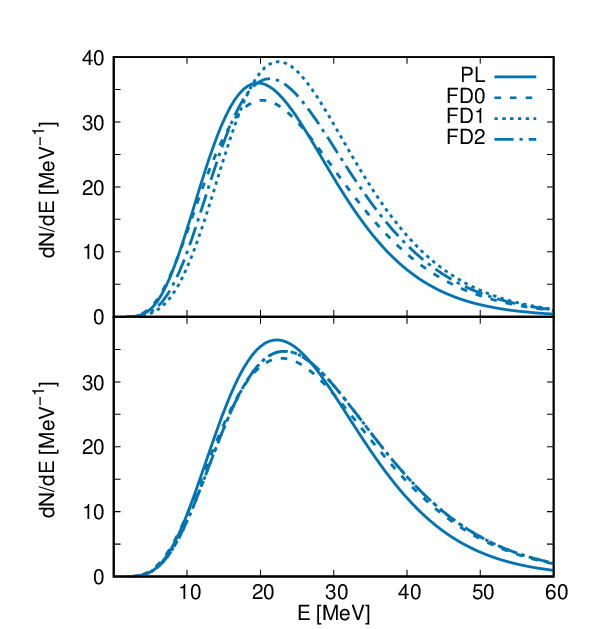}
\end{center}
\caption{Energy distribution of neutrino events, for the active-active neutrino mixing scenario. Top row: normal hierarchy; bottom row: inverse hierarchy. Solid line: PL; dashed line: FD0; dotted line: FD1; dash-dotted line: FD2.} \label{fig:eventos-3act}
\end{figure}

In Fig. \ref{fig:eventos-3act} we present our result for the neutrino event number as a function of the neutrino energy for all the scenarios considered in Fig. \ref{fig:flujo-3act}. As one can see for the normal hierarchy, different distribution functions give different results, however, for the inverse hierarchy, the events are quite similar in all the cases. 
We have also computed the total event number (see Table \ref{N-3act}) for each scenario. As one can see, the normal hierarchy provides lower counts than the inverse hierarchy.  We have computed the total event-number without considering neutrino oscillations, for comparison, and found out that it is smaller than the one obtained in presence of active neutrino oscillations. Also, the power law distribution function gives lower counts than the ones obtained using any of the Fermi Dirac distribution function.
\begin{table}[htp]
\begin{center}
\begin{tabular}{|c|c|c|c|c|}\hline
Distribution function & no/osc. & NH & IH \\ \hline
PL &$760$&$803$&$901$\\ \hline
FD0&$786$&$830$&$930$\\ \hline
FD1&$920$&$930$&$952$\\ \hline
FD2&$856$&$886$&$952$\\ \hline
\end{tabular}
\caption{Calculated number event in a detector for active-active neutrino oscillations, for the different distribution functions discussed in the text. The second column shows the values obtained without taking oscillations between neutrino flavors into account, the third and fourth columns give the values obtained with the normal (NH) and inverse (IH) mass hierarchies, respectively.  } \label{N-3act}
\end{center}
\end{table}

\subsection{Active-sterile neutrino mixing}

\begin{figure}[ht]
\begin{center}
\includegraphics[width=0.5\textwidth]{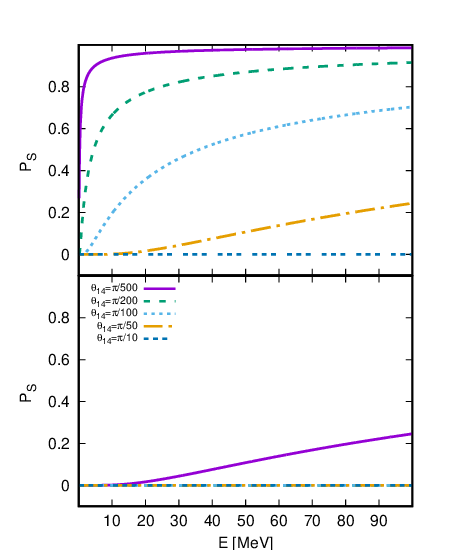}
\end{center}
\caption{Crossing probability for the S resonance as a function of the neutrino energy for different active-sterile neutrino mixing parameters. Top figure: $\Delta m^2_{41}=10^{-3}\, {\rm eV}^2$; bottom figure: $\Delta m^2_{41}=1\, {\rm eV}^2$.}\label{fig:ps-m}
\end{figure}

\begin{figure}[ht]
\begin{center}
\includegraphics[width=0.5\textwidth]{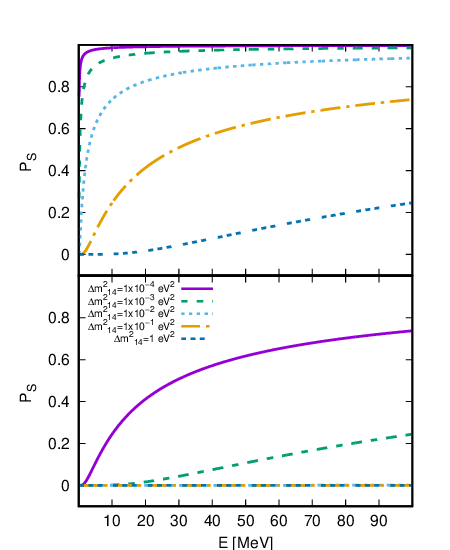}
\end{center}
\caption{Crossing probability for the S resonance as a function of the neutrino energy for different active-sterile neutrino mixing parameters. Top figure: $\theta_{14}=\pi/500$; bottom figure: $\theta_{14}=\pi/50$.}\label{fig:ps-theta}
\end{figure}

We have fixed the SN environment parameters and the active-active neutrino oscillation parameters the same as the previous section and studied the crossing probability for the active-sterile mixing. From Figs. \ref{fig:ps-m} and \ref{fig:ps-theta} one can see that the crossing probability is larger for larger values of the neutrino energy and that it depends on the active-sterile mixing parameters. These results are in complete agreement with the ones obtained by other authors \cite{Dighe:2000,Esmaili:2014}. 

\begin{figure}[ht]
\begin{center}
 \includegraphics[width=0.45\textwidth]{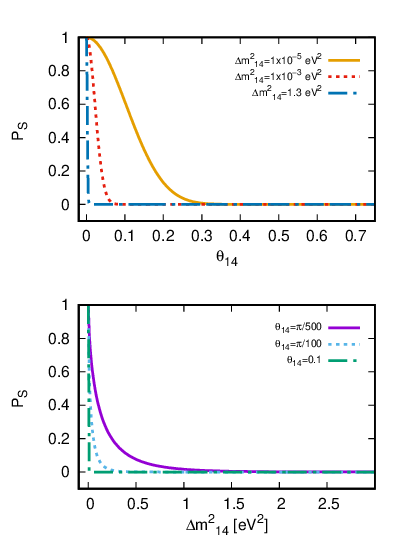}
\end{center}
\caption{Crossing probability for the S resonance at $E=20\, {\rm MeV}$. Top figure: $P_S$ as a function of the mixing angle and for different values of the square mass difference; bottom figure: $P_S$ as a function of $\Delta m^2_{14}$ for different values of $\theta_{14}$ in radians.} \label{fig:ps-e}
\end{figure}

From our results we notice that the adiabatic approximation is quite good for $\Delta m^2_{41} \sim 1 \, {\rm eV}^2$ except for small mixing angles, that is $\theta_{14}<\pi/200$. For smaller square mass difference $\Delta m^2_{41}< 10^{-2}\, {\rm eV}^2$, the adiabatic approximation is not suitable for $\theta_{14}<\pi/50$. If one fixes the neutrino energy, for instance at $E=20 \, {\rm MeV}$ as in Fig. \ref{fig:ps-e}, one can see that the crossing probability is one for zero active-sterile neutrino mixing angle and for very small values of $\Delta m^2_{41}$. 

\begin{figure}[ht]
\begin{center}
\includegraphics[width=0.5\textwidth]{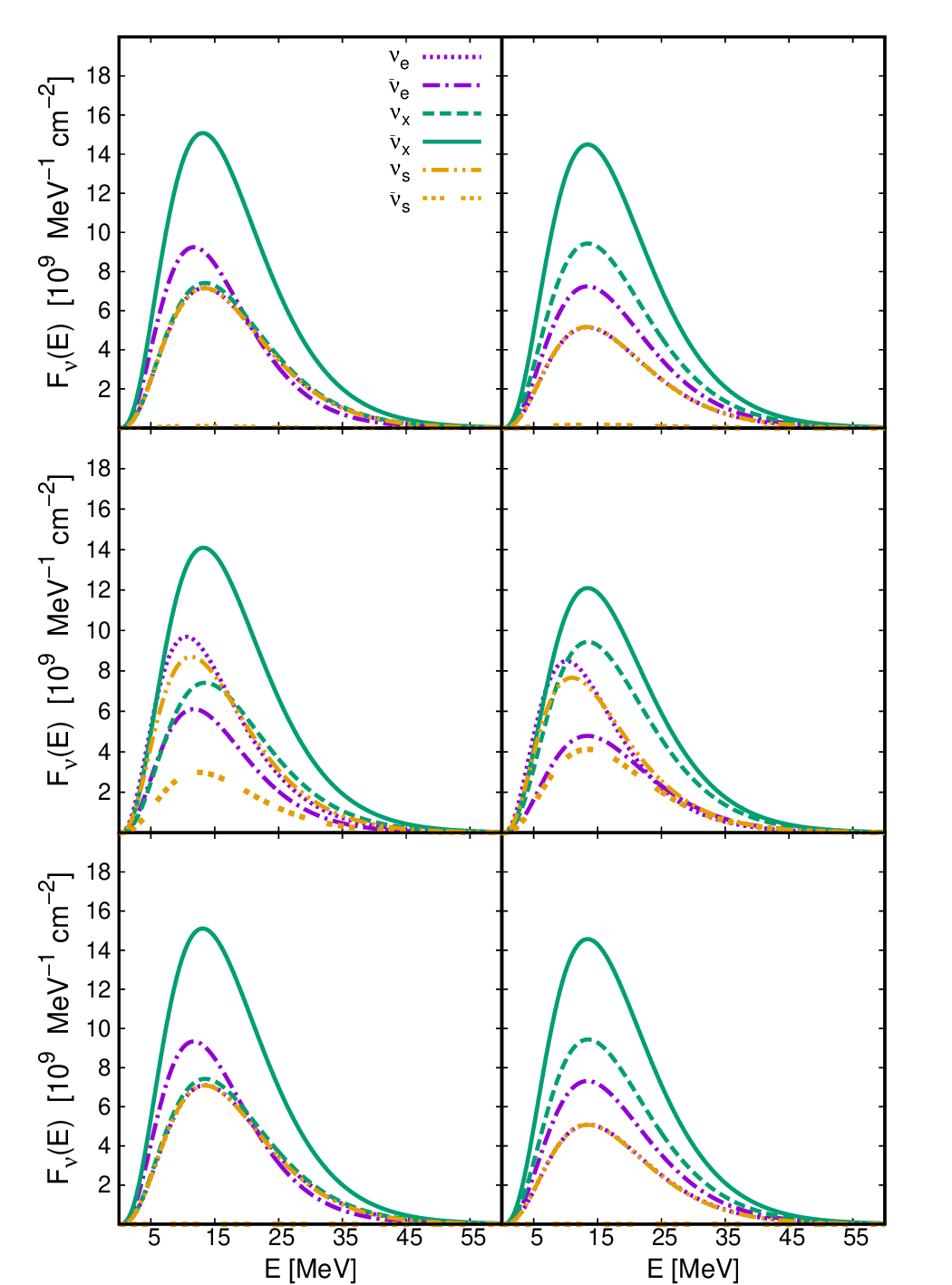}
\end{center}
\caption{Neutrino's fluxes at the detector as a function of the neutrino energy, for the active-sterile neutrino mixing scenario. 
Top row: $\theta_{14}=0.1$; middle row: $\theta_{14}=0.6$; bottom row: $\theta_{14}=0.006$. Left column: normal hierarchy; right column: inverse hierarchy. All the cases $\Delta m^2_{14}= 1.3\,{\rm eV}^2$ and no initial sterile neutrino $\left( F_{\nu_s}^0=0\right)$. The calculation was performed by using a power-law distribution function for active neutrinos.} \label{fig:flujo-est-fs0}
\end{figure}

To compute the SN neutrino fluxes with active-sterile neutrino mixing, we use the values for the mean energies given in the previous section for the active flavors. The active-sterile square mass difference was fixed to $\Delta m^2_{14}= 1.3\,{\rm eV}^2$ \cite{Conrad:2013,Maltoni:2007,Dentler:2018,Boser:2020,Gariazzo:2015,Diaz:2019}. In Fig. \ref{fig:flujo-est-fs0}  we show the SN neutrino fluxes that arrive at the detector as a function of the neutrino energy, in the absence of sterile neutrinos in the SN $\left( F_{\nu_s}^0=0\right)$ and for different mixing angles. In all of the following cases we have considered, for both active and sterile neutrino fluxes, a power-law distribution function. As one can see, the larger is $\theta_{14}$, the lower is the electron antineutrino flux. 

\begin{figure*}[ht]
\includegraphics[width=0.495\linewidth]{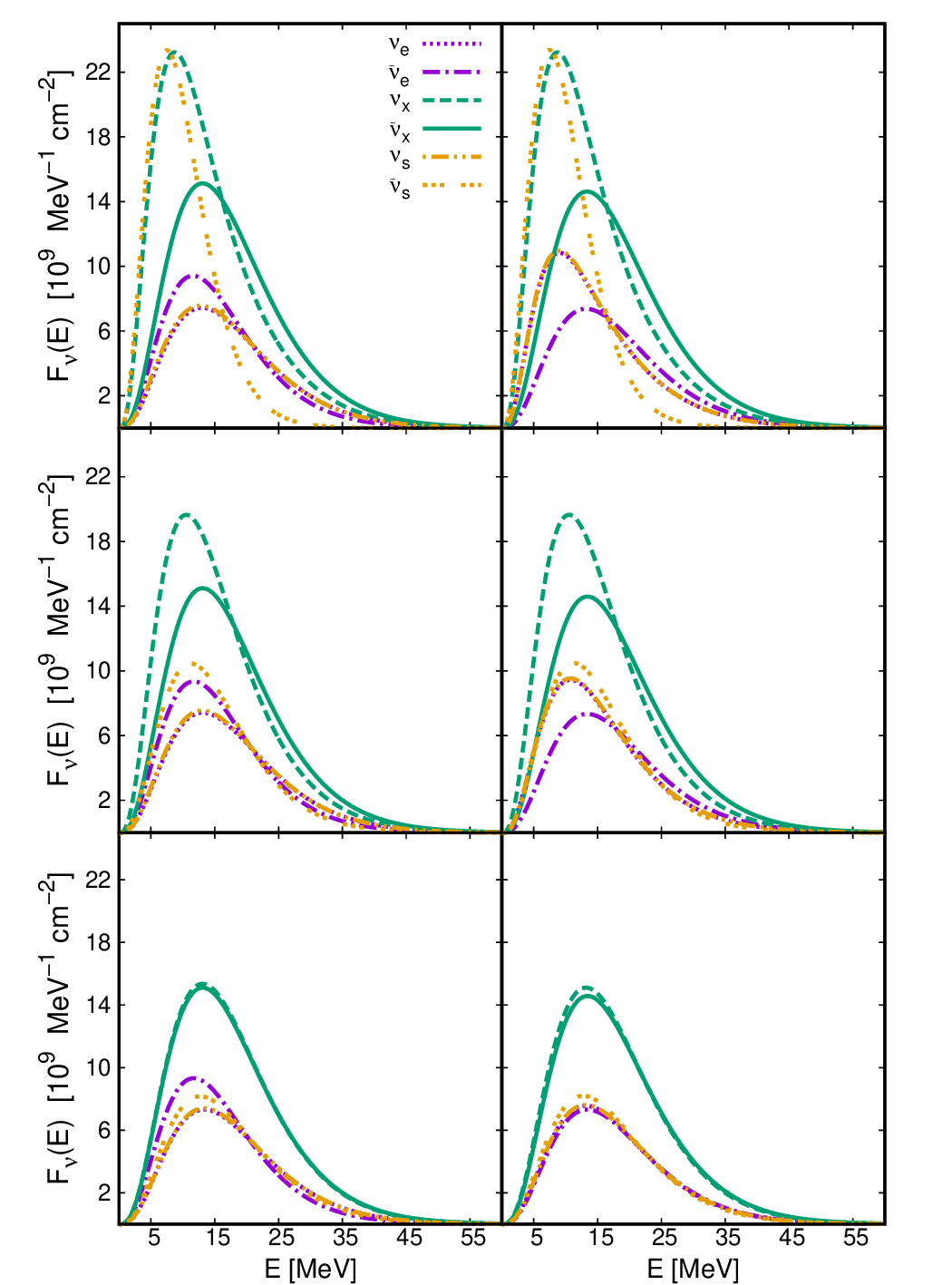}
\includegraphics[width=0.495\linewidth]{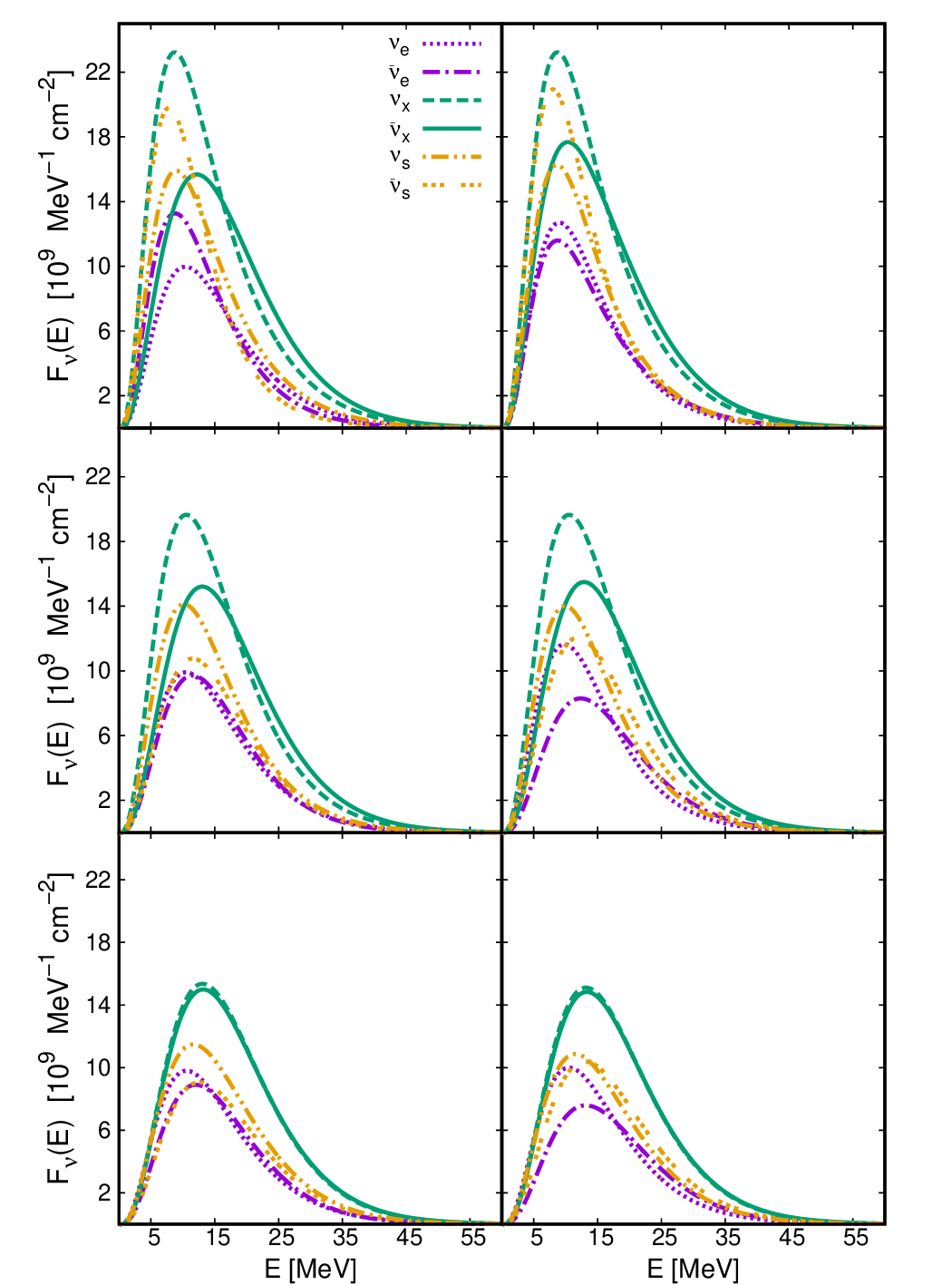}
\caption{Neutrino's fluxes at the detector as a function of the neutrino energy, for the active-sterile neutrino mixing scenario and $\Delta m^2_{14}= 1.3\,{\rm eV}^2$ and with initial sterile neutrino. Top row: $\braket{E_{\nu_s}}=10 \, {\rm MeV}$; middle row: $\braket{E_{\nu_s}}=\braket{E_{\nu_e}}=12 \, {\rm MeV}$; bottom row: $\braket{E_{\nu_s}}=17\, {\rm MeV}$. For each inset: left column: normal hierarchy; right column: inverse hierarchy. The computation was performed by using a power law distribution function for all neutrinos. Left panel: $\theta_{14}=0.1$; right panel: $\theta_{14}=0.6$.} \label{fig:flujo-est-fs-01}
\end{figure*}

When a possible non-vanishing initial flux of sterile neutrinos $\left( F_{\nu_s}^0\neq 0\right)$ is considered, the fluxes are modified according to Fig. \ref{fig:flujo-est-fs-01}. The flux for the electron type antineutrino changes if one changes the active-sterile mixing or the mean energy of the sterile neutrino. The number of events in the detector also would change.

\begin{figure}[ht]
\centering
\includegraphics[width=0.7 \textwidth]{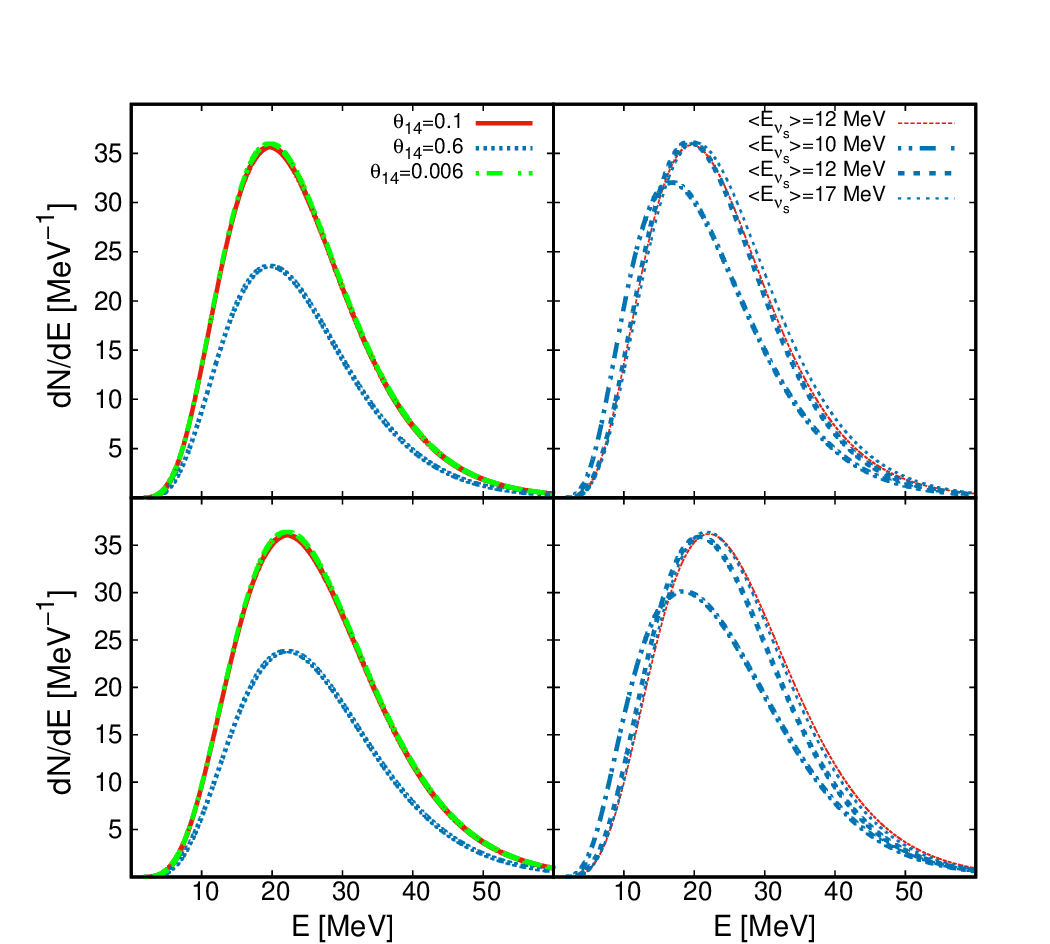}
\caption{Energy distribution of neutrino events, for the active-sterile neutrino mixing scenario. Top row: normal hierarchy; bottom row: inverse hierarchy. Left column: with out sterile neutrino in the SN $\left(F_{\nu_s}^0=0\right)$; right column: with initial sterile-neutrinos flux for several mean energies. Red lines: $\theta_{14}=0.1$; blue lines: $\theta_{14}=0.6$; green lines: $\theta_{14}=0.006$} \label{fig:eventos-est}
\end{figure}

The neutrino number events in the detector as a function of the neutrino energy are presented in Fig. \ref{fig:eventos-est}. In particular, we show the energy distribution of the number event without initial sterile neutrinos and different mixing angles (left column), and for non-vanishing initial flux of sterile neutrinos, for different mean energies and mixing angles (right column). As one can see, the inclusion of a  sterile neutrino with a large mixing angle and a small mean energy affects the distribution of events in the detector, specially the energy at which the maximum number of events is predicted. This is translated into the total number of events (see Table \ref{N-est}).
\begin{table}[ht]
\begin{center}
\begin{tabular}{|c|c|c|c|c|}\hline
$F^0_{\nu_s}$ &$\theta_{14}$ &$\braket{E_{\nu_s}}$ & NH & IH \\ \hline
$0$ & $0.1$ & --& $796$ & $889$  \\ \hline
$0$ & $0.6$ & --& $527$ & $587$ \\ \hline
 & & $10 \, {\rm MeV}$ & $801$ & $894$ \\ \cline{3-5}
$\neq 0$ & $0.1$ &$12 \, {\rm MeV}$ & $804$ & $897$ \\ \cline{3-5}
& & $17 \, {\rm MeV}$ & $805$ & $898$ \\ \hline
 & & $10 \, {\rm MeV}$ & $699$ & $760$ \\ \cline{3-5}
$\neq 0$ & $0.6$ &$12 \, {\rm MeV}$ & $789$ & $850$ \\ \cline{3-5}
& & $17 \, {\rm MeV}$ & $822$ & $883$ \\ \hline
\end{tabular}
\caption{Calculated number event in a scintillator detector for active-sterile neutrino oscillation.} \label{N-est}
\end{center}
\end{table}

\begin{figure}[ht]
\begin{center}
\includegraphics[width=0.45 \textwidth]{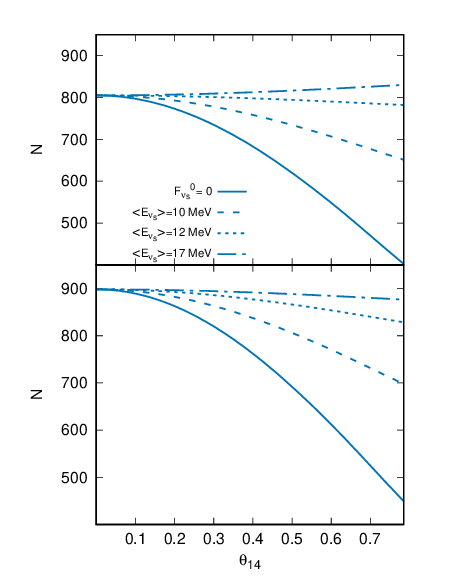}
\end{center}
\caption{Neutrino number events as a function of the active-sterile neutrino mixing angle. Top row: normal hierarchy; bottom row: inverse hierarchy. Solid line: no sterile neutrino in the SN; small dashed line: $\braket{E_{\nu_s}}=10 \, {\rm MeV}$; dotted line: $\braket{E_{\nu_s}}=12 \, {\rm MeV}$; long dashed line: $\braket{E_{\nu_s}}=17\, {\rm MeV}$. In all cases $\Delta m^2_{14}= 1.3\,{\rm eV}^2$. } \label{fig:numero-est}
\end{figure}

In Fig. \ref{fig:numero-est} we show the number of events in the detector as a function of the active-sterile mixing angle. We have found that the number of events does not depend on the square mass difference, but strongly depends on the mixing angle. In all the studied cases, the neutrino inverse hierarchy produces more events than the normal hierarchy. The number of events for $F^0_{\nu_s}=0$ decreases for larger mixing angles. If we consider sterile neutrinos inside the supernova, the mean energy of these neutrinos affects the value of $N$. For $F^0_{\nu_s}\neq 0$ the value of $N$ also decreases with the mixing angle except for NH with sterile neutrino mean energy larger than the electron neutrino mean energy.

\subsection{Departure from Poisson's statistics}

Since the data from SN1987A do not allow to study the neutrino energy distribution in detail \cite{Kam:1987,IMB:1987,Baskan:1987} and there is a lack of a quantitative estimation of theoretical uncertainties, it is worth exploring deviations of our theoretical results respect to the data that come from a counting experiment wich follows a Poisson statistics
\begin{equation}
    P(k;\mu)=\frac{\mu^k}{k!}\exp{(-\mu)}
\end{equation}
where $\mu$ is the average number of occurrences in a given interval and $k$ is the discrete random variable. For $\mu$ we have adopted the value corresponding to the 3 active neutrino mixing scenario following Ref. \cite{Tang:2020}. The statistical indicator for Poisson distributions is \cite{Mighell:1999,BAKER:1984}
\begin{equation}\label{chi}
    \Delta \chi^2(\theta_{14},\Delta m^2_{41})=2\sum_{i=1}^N \left( m_i-n_i+n_i\ln{\left(\frac{n_i}{m_i}\right)}\right) 
\end{equation}
The sum runs over energy bins, $m_i$ are the theoretical values for the number of events for the i-th bin, which depend on  $\theta_{14}$ and $\Delta m^2_{41}$ and $n_i$ are the number of events calculated with the Poisson distribution for each energy bin with an error $\sigma_i=\sqrt{n_i}$. Following Refs. \cite{Araki:2005,Moller:2018} we have studied the number of events along 32 bins of $2 \, \rm{MeV}$ width. We have performed the analysis for an active-sterile mixing angle in the range $0 \leq \theta_{14}\leq \pi/4$, and fixed square mass difference  $\Delta m^2_{41}= 1.3\,{\rm eV}^2$. The best value of the mixing angle is determined by minimizing Eq. (\ref{chi}). For all the studied cases, we have obtained a best fit value for $\theta_{14}$ shown in Table \ref{ajuste}.  
\begin{table}[ht]
\begin{center}
\begin{tabular}{|c|c|c|c|c|}\hline
Hierarchy &$F^0_{\nu_s}$ &$\braket{E_{\nu_s}}$& $\theta_{14}\pm \sigma$ & $\frac{\Delta \chi^2}{N-1}$ \\ \hline
 NH&0 & -- & $0.044 \pm 0.021$& 1.32 \\ \cline{2-5}
 & &10 MeV & $0.032 \pm 0.016$ & 1.32  \\  \cline{3-5}
  &$\neq0 $ &12 MeV & $0.033 \pm 0.015$  &1.32 \\ \cline{3-5}
 & &17 MeV & $0.044 \pm 0.025$ &1.30 \\ \hline
  IH&0 & -- & $0.016 \pm 0.009$& 1.32\\ \cline{2-5}
 & &10 MeV & $0.016\pm0.009$ & 1.37\\  \cline{3-5}
  &$\neq0 $ &12 MeV & $0.016\pm0.009$  & 1.32 \\ \cline{3-5}
 & &17 MeV & $0.016\pm0.010$ & 1.37\\ \hline
\end{tabular}
\caption{Best-fit values of the mixing angle $\theta_{14}$ in radians at $1\sigma$. The reduced $\chi$-square value of each of the fits are given in the last column.} \label{ajuste}
\end{center}
\end{table}

Although the analysis carried out indicates a preference for the $3 + 1$ model with a small but non-zero angle $\theta_{14}$, it should be noted that all cases are consistent with $ \theta_{14}= 0 $ at 3$\sigma$. This indicates that the data can be represented by considering only active-active neutrino oscillations in the formalism or by the inclusion of active-sterile neutrino mixing with a small mixing angle.

So far, the data and the analysis presented in this section are those provided by measurements coming from liquid sctinllators. To complete the picture about the mixing with $\nu_s$ it would be necessary to perform a similar analysis for data gathered by different  types of neutrino detectors and complement this kind of study with the determination of the  neutrino spectral shape.

\section{Conclusions}
\label{conclusiones}

In this work, we have included massive neutrinos, neutrino oscillation, and a light sterile neutrino in the formalism of SN neutrino fluxes and their detection in liquid scintillators through inverse beta decay reactions. 

We have computed the neutrino fluxes and the number of detected events with active neutrino mixing. We have found  that the neutrino fluxes are sensitive to differences between the normal and inverse mass hierarchies. The power-law distribution generates a lower number of total events than the Fermi-Dirac distribution function. Furthermore, the number of expected events increases with respect to the case for which oscillations are not included.

Then, we have computed the crossing probabilities, neutrino fluxes and the number of detected events as a function of the mixing parameters for the active-sterile sector. As the energy increases, the crossing of the resonances becomes less adiabatic, and when both mixing parameters are small, the non-adiabaticity is more pronounced. We have found that the adiabatic approximation for $P_S$ is no longer valid for $\Delta m^2_{41}\sim 1 \rm{eV}^2$ and $0<\theta_{14} <\pi/200$ and for $\Delta m^2_{41} <1 \times 10^{-2}\, \rm{eV}^2$ with $\theta_{14} <\pi/50$.

When there is not initial sterile neutrino flux, that is $F^0_{\nu_s}=0$, the number of events decreases for large mixing angles $\theta_{14}$. However, for a non-zero initial sterile neutrino flux, the number of events depends on the mean energy used to calculate the spectral function  $F^0_{\nu_s}$. This event detection does not depend on the mass-square difference (since the flux ${\bar{\nu}_e}$ does not depend on $P_S$). In all the studied cases, the IH produces a larger number of events than the NH. 

Finally, we have performed a statistical test to set limits on the mixing angle between active and sterile neutrinos. We have found that in all the cases the best value of the active-sterile mixing angle is small but not null. Studying neutrinos from SN in different types of detectors can be useful to provide a definitive conclusion about the existence of eV-scale sterile neutrinos and the preferred mixing scheme.

We hope that a better understanding of the expected spectrum of events and fluxes will help to the analysis of the next supernova event in our galaxy, as well as to set prospects for the construction of future detectors, as the one planned to be hosted by the ANDES lab \cite{Machado:2012,Civitarese:2015,Bertou:2012}.
\section*{Acknowledgements}
This work was supported by a grant (PIP-616) of the National Research Council of Argentina (CONICET), and by a research-grant (PICT 140492) of the National Agency for the Promotion of Science and Technology (ANPCYT) of Argentina. O. C. and M. E. M. are members of the Scientific Research Career of the CONICET, M. M. S. is a Post Doctoral fellow of the CONICET.

\bibliographystyle{ws-ijmpe}
\bibliography{biblio}

\begin{thebibliography}{10}

\bibitem{Woosley:1986}
S.~E. {Woosley} and T.~A. {Weaver}, {\em Annu. Rev. Astron. Astrophys.} {\bf
  24}  (1986) 205.

\bibitem{Buras:2003}
R.~Buras {\em et~al.}, {\em Astrophys. J.} {\bf 587}  (2003) 320.

\bibitem{Janka:2016}
H.-T. Janka, Neutrino emission from supernovae, in {\em Handbook of
  Supernovae\/},  eds. A.~W. Alsabti and P.~Murdin (Springer International
  Publishing, Cham, 2016), Cham, pp. 1--30.

\bibitem{Balantekin:2005}
A.~B. Balantekin and H.~Yüksel, {\em New J. Phys.} {\bf 7}  (2005) 51.

\bibitem{Tamborra:2012}
I.~Tamborra, G.~G. Raffelt, L.~Hudepohl and H.-T. Janka, {\em J. Cosmol.
  Astropart. Phys.} {\bf 01}  (2012)   013.

\bibitem{Rafflet:2007}
G.~G. Raffelt and A.~Y. Smirnov, {\em Phys. Rev. D} {\bf 76}  (2007)
  081301(R).

\bibitem{Dasgupta:2009}
B.~{Dasgupta}, A.~{Dighe}, G.~G. {Raffelt} and A.~Y. {Smirnov}, {\em Phys. Rev.
  Lett.} {\bf 103}  (2009)   051105,
  \href{http://arxiv.org/abs/0904.3542}{{\ttfamily arXiv:0904.3542 [hep-ph]}}.

\bibitem{Giunti:2003}
C.~{Giunti} and M.~{Laveder}, {\em arXiv e-prints}  (October 2003)
  \href{http://arxiv.org/abs/hep-ph/0310238}{{\ttfamily arXiv:hep-ph/0310238}}.

\bibitem{Abe:2016}
 Super-Kamiokande Collaboration Collaboration (K.~{Abe} {\em et~al.}), {\em
  Phys. Rev. D} {\bf 94} (Sep 2016)   052010.

\bibitem{Gallex:1998}
 GALLEX Collaboration (W.~Hampel {\em et~al.}), {\em Phys. Lett. B} {\bf 447}
  (1999) 127.

\bibitem{Kajita:2010}
T.~Kajita, {\em Proc. Japan Acad. B} {\bf 86}  (2010) 303.

\bibitem{Lunardini:2001}
C.~Lunardini and A.~Y. Smirnov, {\em Phys. Rev. D} {\bf 63}  (2001)   073009.

\bibitem{Hirata:1988}
K.~S. Hirata, T.~Kajita, M.~Koshiba, M.~Nakahata, Y.~Oyama, N.~Sato {\em
  et~al.}, {\em Phys. Rev. D} {\bf 38} (Jul 1988) 448.

\bibitem{Qian:2018}
Y.~{Qian}, {\em Sci. China Phiys. Mech.} {\bf 61}  (2018)   49501,
  \href{http://arxiv.org/abs/1801.09554}{{\ttfamily arXiv:1801.09554}}.

\bibitem{Mirizzi:2016}
A.~{Mirizzi} {\em et~al.}, {\em Nuovo Cimento Rivista Serie} {\bf 39}  (2016)
  1, \href{http://arxiv.org/abs/1508.00785}{{\ttfamily arXiv:1508.00785}}.

\bibitem{Takahashi:2001}
K.~Takahashi, M.~Watanabe, K.~Sato and T.~Totani, {\em Phys. Rev. D} {\bf 64}
  (2001)   093004, \href{http://arxiv.org/abs/hep-ph/0105204}{{\ttfamily
  arXiv:hep-ph/0105204}}.

\bibitem{Machado:2012}
P.~A.~N. Machado, T.~Muhlbeier, H.~Nunokawa and R.~Z. Funchal, {\em Phys. Rev.
  D} {\bf 86}  (2012)   125001,
  \href{http://arxiv.org/abs/1207.5454}{{\ttfamily arXiv:1207.5454}}.

\bibitem{Aguilar-Arevalo:2018}
 MiniBooNE Collaboration Collaboration (A.~Arevalo, B.~Brown, L.~Bugel,
  G.~Cheng, J.~Conrad and R.~Cooper), {\em Phys. Rev. Lett.} {\bf 121}  (2018)
   221801.

\bibitem{Dentler:2018}
M.~{Dentler} {\em et~al.}, {\em J. High Energy Phys.} {\bf 2018}  (2018)  ~10,
  \href{http://arxiv.org/abs/1803.10661}{{\ttfamily arXiv:1803.10661
  [hep-ph]}}.

\bibitem{Mention11}
G.~Mention, M.~Fechner, T.~Lasserre, T.~A. Mueller, D.~Lhuillier, M.~Cribier
  and A.~Letourneau, {\em Phys. Rev. D} {\bf 83}  (2011)   073006.

\bibitem{Acero:2007}
M.~A. Acero, C.~Giunti and M.~Laveder, {\em Phys. Rev. D} {\bf 78}  (2008)
  073009, \href{http://arxiv.org/abs/0711.4222}{{\ttfamily arXiv:0711.4222
  [hep-ph]}}.

\bibitem{Giunti:2011}
C.~Giunti and M.~Laveder, {\em Phys. Rev. D} {\bf 84}  (2011)   073008.

\bibitem{Kopp:2013}
J.~Kopp, P.~A.~N. Machado, M.~Maltoni and T.~Schwetz, {\em J. High Energy
  Phys.} {\bf 05}  (2013)  ~50.

\bibitem{Boyarsky:2009}
A.~{Boyarsky}, O.~{Ruchayskiy} and M.~{Shaposhnikov}, {\em Annu. Rev. Nucl.
  Part. Sci.} {\bf 59}  (2009) 191,
  \href{http://arxiv.org/abs/0901.0011}{{\ttfamily arXiv:0901.0011 [hep-ph]}}.

\bibitem{Mohapatra:2004}
R.~N. Mohapatra and P.~B. Pal, {\em Massive Neutrinos in Physics and
  Astrophysics}, 3rd edn. (WORLD SCIENTIFIC, 2004).

\bibitem{Mclaughlin:1999}
G.~C. McLaughlin, J.~M. Fetter, A.~B. Balantekin and G.~M. Fuller, {\em Phys.
  Rev. C} {\bf 59} (May 1999) 2873.

\bibitem{Caldwell:2000}
D.~O. Caldwell, G.~M. Fuller and Y.-Z. Qian, {\em Phys. Rev. D} {\bf 61} (May
  2000)   123005.

\bibitem{Fetter:2003}
J.~Fetter, G.~McLaughlin, A.~Balantekin and G.~Fuller, {\em Astropart.Phys.}
  {\bf 18}  (2003) 433 .

\bibitem{Saez:2018}
M.~M. {Saez}, O.~{Civitarese} and M.~E. {Mosquera}, {\em Int. J. Mod. Phys. D}
  {\bf 27}  (2018) 1850116, \href{http://arxiv.org/abs/1808.03249}{{\ttfamily
  arXiv:1808.03249 [hep-ph]}}.

\bibitem{Wu:2014}
M.~R. Wu, T.~Fischer, L.~Huther, G.~Martinez-Pinedo and Y.~Z. Qian, {\em Phys.
  Rev. D} {\bf 89}  (2014)   061303(R),
  \href{http://arxiv.org/abs/1305.2382}{{\ttfamily arXiv:1305.2382
  [astro-ph.HE]}}.

\bibitem{Esmaili:2014}
A.~{Esmaili}, O.~L.~G. {Peres} and P.~D. {Serpico}, {\em Phys. Rev. D} {\bf 90}
   (2014)   033013, \href{http://arxiv.org/abs/1402.1453}{{\ttfamily
  arXiv:1402.1453 [hep-ph]}}.

\bibitem{Tang:2020}
J.~Tang, T.~Wang and M.-R. Wu, {\em J. Cosmol. Astropart. Phys.} {\bf 2020}
  (Oct 2020) 038.

\bibitem{SNOplus:2015}
 SNO+ Collaboration (L.~Sibley), {\em AIP Conf. Proc.} {\bf 1604}  (2015) 449.

\bibitem{Suzuki:2006}
T.~Suzuki, S.~Chiba, T.~Yoshida, T.~Kajino and T.~Otsuka, {\em Physical Review
  C} {\bf 74}  (2006)   034307.

\bibitem{Suzuki:2019}
T.~Suzuki, A.~B. Balantekin, T.~Kajino and S.~Chiba, {\em Journal of Physics G}
  {\bf 46}  (2019).

\bibitem{Choubey:2006}
S.~Choubey, N.~P. Harries and G.~G. Ross, {\em Phys. Rev. D} {\bf 74} (Sep
  2006)   053010.

\bibitem{Huang:2015}
M.-Y. Huang, X.-H. Guo and B.-L. Young, {\em Chin. Phys. C} {\bf 40}  (2016)
  073102, \href{http://arxiv.org/abs/1511.00806}{{\ttfamily arXiv:1511.00806
  [hep-ph]}}.

\bibitem{Hirata:1987}
K.~Hirata, T.~Kajita, M.~Koshiba, M.~Nakahata, Y.~Oyama and N.~Sato, {\em Phys.
  Rev. Lett.} {\bf 58}  (1987) 1490.

\bibitem{Bionta:1987}
R.~M. Bionta, G.~Blewitt, C.~B. Bratton, D.~Casper, A.~Ciocio and R.~Claus,
  {\em Phys. Rev. Lett.} {\bf 58}  (1987) 1494.

\bibitem{Keil:2003}
M.~T. Keil, G.~G. Raffelt and H.-T. Janka, {\em Astrophys. J.} {\bf 590}
  (2003) 971, \href{http://arxiv.org/abs/astro-ph/0208035}{{\ttfamily
  arXiv:astro-ph/0208035}}.

\bibitem{Janka:1989}
H.~T. {Janka} and W.~{Hillebrandt}, {\em Astron. Astrophys.} {\bf 224}  (1989)
  49.

\bibitem{Lang:2016}
R.~F. Lang, C.~McCabe, S.~Reichard, M.~Selvi and I.~Tamborra, {\em Phys. Rev.
  D} {\bf 94}  (2016)   103009,
  \href{http://arxiv.org/abs/1606.09243}{{\ttfamily arXiv:1606.09243
  [astro-ph.HE]}}.

\bibitem{Duan:2010b}
H.~Duan, G.~M. Fuller and Y.-Z. Qian, {\em Ann. Rev. Nucl. Part. Sci.} {\bf 60}
   (2010) 569, \href{http://arxiv.org/abs/1001.2799}{{\ttfamily arXiv:1001.2799
  [hep-ph]}}.

\bibitem{Chakraborty:2011}
S.~Chakraborty, T.~Fischer, A.~Mirizzi, N.~Saviano and R.~Tomas, {\em Phys.
  Rev. Lett.} {\bf 107}  (2011)   151101,
  \href{http://arxiv.org/abs/1104.4031}{{\ttfamily arXiv:1104.4031 [hep-ph]}}.

\bibitem{Esteban:2008}
A.~Esteban-Pretel, A.~Mirizzi, S.~Pastor, R.~Tomas, G.~G. Raffelt, P.~D.
  Serpico and G.~Sigl, {\em Phys. Rev. D} {\bf 78}  (2008)   085012,
  \href{http://arxiv.org/abs/0807.0659}{{\ttfamily arXiv:0807.0659
  [astro-ph]}}.

\bibitem{Maki:1962}
Z.~Maki, M.~Nakagawa and S.~Sakata, {\em Prog. Theor. Phys.} {\bf 28}  (1962)
  870.

\bibitem{Giganti:2018}
C.~{Giganti}, S.~{Lavignac} and M.~{Zito}, {\em Prog. Part. Nucl. Phys.} {\bf
  98}  (2018) 1, \href{http://arxiv.org/abs/1710.00715}{{\ttfamily
  arXiv:1710.00715 [hep-ex]}}.

\bibitem{Gariazzo:2017}
S.~{Gariazzo}, C.~{Giunti}, M.~{Laveder} and Y.~F. {Li}, {\em J. High Energy
  Phys.} {\bf 06}  (2017)   135,
  \href{http://arxiv.org/abs/1703.00860}{{\ttfamily arXiv:1703.00860
  [hep-ph]}}.

\bibitem{Collin:2016}
G.~H. {Collin}, C.~A. {Arg{\"u}elles}, J.~M. {Conrad} and M.~H. {Shaevitz},
  {\em Phys. Rev. Lett.} {\bf 117}  (2016)   221801,
  \href{http://arxiv.org/abs/1607.00011}{{\ttfamily arXiv:1607.00011
  [hep-ph]}}.

\bibitem{Dighe:2000}
A.~S. Dighe and A.~Y. Smirnov, {\em Phys. Rev. D} {\bf 62}  (2000)   033007,
  \href{http://arxiv.org/abs/hep-ph/9907423}{{\ttfamily arXiv:hep-ph/9907423}}.

\bibitem{Smirnov:2003}
A.~Y. {Smirnov}, {\em arXiv e-prints}   (2003)
  \href{http://arxiv.org/abs/hep-ph/0305106}{{\ttfamily arXiv:hep-ph/0305106
  [hep-ph]}}.

\bibitem{Landau:1932}
L.~D. Landau, {\em Z. Sowjetunion} {\bf 2}  (1932) 46.

\bibitem{Zener:1932}
C.~Zener and R.~H. Fowler, {\em Proc. R. Soc. London} {\bf 137}  (1932) 696.

\bibitem{Fogli:2002}
G.~L. {Fogli}, E.~{Lisi}, D.~{Montanino} and A.~{Palazzo}, {\em Phys. Rev. D}
  {\bf 65}  (2002)   073008,
  \href{http://arxiv.org/abs/hep-ph/0111199}{{\ttfamily arXiv:hep-ph/0111199}}.

\bibitem{Brown:1982}
G.~Brown, H.~Bethe and G.~Baym, {\em Nucl. Phys. A} {\bf 375}  (1982) 481 .

\bibitem{Mikheev:1986}
S.~Mikheyev and A.~Smirnov, {\em Sov. J. Nucl. Phys.} {\bf 42}  (1985) 913.

\bibitem{Altmann:2005}
M.~Altmann {\em et~al.}, {\em Phys. Lett. B} {\bf 616}  (2005) 174 .

\bibitem{Sage:1999}
 SAGE Collaboration Collaboration (J.~N. Abdurashitov, V.~N. Gavrin, S.~V.
  Girin, V.~V. Gorbachev, T.~V. Ibragimova and A.~V. Kalikhov), {\em Phys. Rev.
  C} {\bf 60} (Oct 1999)   055801.

\bibitem{Nakamura:1994}
K.~Nakamura, T.~Kajita, M.~Nakahata and A.~Suzuki, Kamiokande, in {\em Physics
  and Astrophysics of Neutrinos\/},  ed. S.~A. Fukugita~M. (Springer
  International Publishing, 1994), pp. 249--387.

\bibitem{Fukuda:2002}
 Super-Kamiokande Collaboration (Y.~Fukuda {\em et~al.}), {\em Nucl. Instrum.
  Meth. A} {\bf 501}  (2003) 418.

\bibitem{Kamland:2005}
T.~{Araki} {\em et~al.}, {\em Phys. Rev. Lett.} {\bf 94}  (2005)   081801,
  \href{http://arxiv.org/abs/hep-ex/0406035}{{\ttfamily arXiv:hep-ex/0406035
  [hep-ex]}}.

\bibitem{Borexino:2002}
 Borexino Collaboration Collaboration (G.~{Alimonti} {\em et~al.}), {\em
  Astropart.Phys.} {\bf 16}  (2002) 205,
  \href{http://arxiv.org/abs/hep-ex/0012030}{{\ttfamily arXiv:hep-ex/0012030
  [hep-ex]}}.

\bibitem{LSND:1996}
 LSND Collaboration Collaboration (C.~Athanassopoulos, L.~B. Auerbach, R.~L.
  Burman, I.~Cohen, D.~O. Caldwell and B.~D. Dieterle), {\em Phys. Rev. Lett.}
  {\bf 77}  (1996) 3082.

\bibitem{Strumia:2003}
A.~{Strumia} and F.~{Vissani}, {\em Phys. Lett. B} {\bf 564}  (2003) 42,
  \href{http://arxiv.org/abs/astro-ph/0302055}{{\ttfamily
  arXiv:astro-ph/0302055 [astro-ph]}}.

\bibitem{Cadonati:2002}
L.~Cadonati, F.~Calaprice and M.~Chen, {\em Astropart.Phys.} {\bf 16}  (2002)
  361 .

\bibitem{Civitarese:2015}
O.~Civitarese, { The andes underground laboratory project}, in {\em X Latin
  American Symposium of High Energy Physics\/},  {\em Nucl. Part. Phys. Proc.}
  {\bf \textbf{267-269}} (2015), pp. 377--381.

\bibitem{Mirizzi:2006}
A.~{Mirizzi}, G.~G. {Raffelt} and P.~D. {Serpico}, {\em J. Cosmol. Astropart.
  Phys.} {\bf 05}  (2006)   012,
  \href{http://arxiv.org/abs/astro-ph/0604300}{{\ttfamily
  arXiv:astro-ph/0604300 [astro-ph]}}.

\bibitem{Pdg:2019}
 Particle Data Group Collaboration (M.~Tanabashi {\em et~al.}), {\em Phys. Rev.
  D} {\bf 98}  (2018)   030001.

\bibitem{Conrad:2013}
J.~M. Conrad, W.~C. Louis and M.~H. Shaevitz, {\em Ann. Rev. Nucl. Part. Sci.}
  {\bf 63}  (2013) 45, \href{http://arxiv.org/abs/1306.6494}{{\ttfamily
  arXiv:1306.6494 [hep-ex]}}.

\bibitem{Maltoni:2007}
M.~{Maltoni} and T.~{Schwetz}, {\em Phys. Rev. D.} {\bf 76}  (2007)   093005,
  \href{http://arxiv.org/abs/0705.0107}{{\ttfamily arXiv:0705.0107 [hep-ph]}}.

\bibitem{Boser:2020}
S.~{B{\"o}ser} {\em et~al.}, {\em Prog. Part. Nucl. Phys.} {\bf 111}  (2020)
  103736, \href{http://arxiv.org/abs/1906.01739}{{\ttfamily arXiv:1906.01739
  [hep-ex]}}.

\bibitem{Gariazzo:2015}
S.~Gariazzo {\em et~al.}, {\em J. Phys. G} {\bf 43}  (2015)   033001.

\bibitem{Diaz:2019}
A.~Diaz {\em et~al.}, {\em Physics Reports} {\bf 884}  (2020) 1.

\bibitem{Kam:1987}
R.~{Schaeffer}, Y.~{Declais} and S.~{Jullian}, {\em Nature} {\bf 330} (November
  1987) 142.

\bibitem{IMB:1987}
R.~{Svoboda} {\em et~al.}, { {Neutrinos from Supernova 1987A in the IMB
  Detector}}, in {\em European Southern Observatory Conference and Workshop
  Proceedings\/},  (January 1987), p. 229.

\bibitem{Baskan:1987}
E.~N. {Alekseev} {\em et~al.}, { {Detection of the Neutrino Signal from
  Supernova 1987A Using the INR Baksan Underground Scintillation Telescope}},
  in {\em European Southern Observatory Conference and Workshop Proceedings\/},
   (January 1987), p. 237.

\bibitem{Mighell:1999}
K.~J. Mighell, {\em Astrophys. J.} {\bf 518} (Jun 1999) 380.

\bibitem{BAKER:1984}
S.~Baker and R.~D. Cousins, {\em Nucl. Instrum. Methods} {\bf 221}  (1984) 437.

\bibitem{Araki:2005}
 KamLAND Collaboration Collaboration (T.~e.~a. Araki), {\em Phys. Rev. Lett.}
  {\bf 94} (Mar 2005)   081801.

\bibitem{Moller:2018}
K.~{M{\o}ller}, A.~M. {Suliga}, I.~{Tamborra} and P.~B. {Denton}, {\em J.
  Cosmol. Astropart. Phys.} {\bf 2018} (May 2018)   066,
  \href{http://arxiv.org/abs/1804.03157}{{\ttfamily arXiv:1804.03157
  [astro-ph.HE]}}.

\bibitem{Bertou:2012}
X.~Bertou, {\em Eur. Phys. J. Plus} {\bf 127}  (2012)   104.

\end{thebibliography}
\end{document}